\documentclass[aps,prl,preprint,superscriptaddress,nofootinbib,amsmath,amssymb,longbibliography]{revtex4-2}
\usepackage{adjustbox}
\usepackage[colorlinks]{hyperref} 
\usepackage[normalem]{ulem}

\begin{document}

\title{Unraveling the anomaly in the production of $^{60}$Fe nucleus in massive stars}
\author{Samapti Lakshan}
\affiliation{Department of Physics, Bankura University, Bankura, West Bengal 722155, India}
\author{Le Tan Phuc}
\affiliation{Institute of Fundamental and Applied Sciences, Duy Tan University, Ho Chi Minh City 70000, Vietnam}
\affiliation{Faculty of Natural Sciences, Duy Tan University, Da Nang City 55000, Vietnam}
\author{Deepak Pandit}
\affiliation{Variable Energy Cyclotron Centre, 1/AF-Bidhannagar, Kolkata 700064, India}
\affiliation{Homi Bhabha National Institute, Training School Complex, Anushaktinagar, Mumbai 400094, India}
\author{Srijit Bhattacharya}
\affiliation{Department of Physics, Haldia Goverment College, Purba Medinipur, West Bengal 721657, India}
\author{Le Thi Quynh Huong}
\affiliation{University of Khanh Hoa, Nha Trang City, Khanh Hoa Province 65000, Vietnam}
\author{Nguyen Dinh Dang}
\affiliation{Nuclear Many-Body Theory Laboratory, RIKEN Nishina Center for Accelerator-Based Science,
2-1 Hirosawa, Wako City, 351-0198 Saitama, Japan}
\author{Balaram Dey}
\email{Contact author: dey.balaram@gmail.com}
\affiliation{Department of Physics, Bankura University, Bankura, West Bengal 722155, India}
\author{Nguyen Ngoc Anh}
\email{Contact author: anh.nguyenngoc1@phenikaa-uni.edu.vn}
\affiliation{Phenikaa Institute for Advance Study (PIAS), PHENIKAA University, Hanoi, 12116, Vietnam}
\author{Nguyen Quang Hung}
\email{Contact author: nguyenquanghung5@duytan.edu.vn}
\affiliation{Institute of Fundamental and Applied Sciences, Duy Tan University, Ho Chi Minh City 70000, Vietnam}
\affiliation{Faculty of Natural Sciences, Duy Tan University, Da Nang City 55000, Vietnam}

\date{\today}

\begin{abstract}  

The production of $^{60}$Fe is crucial for nucleosynthesis in massive stars and supernovae. In this work, by using the microscopic EP+IPM (exact pairing plus the independent-particle model) for the nuclear level density (NLD) and extended EP+PDM (exact pairing plus phonon damping model) for the $\gamma$-ray strength function (gSF), we re-evaluate the substantial enhancement of $^{60}$Fe production recently reported in {\it A. Spyrou et al.,  Nat. Comm. {\bf 15}, 9608 (2024)}, which was attributed to an unexpectedly large Maxwellian-averaged cross section (MACS). Our analysis demonstrates that this enhancement indeed originates from the choice of NLD, which, despite being constrained to reproduce the total NLD and gSF data, lacks a reliable spin dependence, a critical input for Hauser-Feshbach calculations of nuclear reaction rate. In contrast, our predictions yield a significantly lower MACS, calling the claimed enhancement into question. In particular, our approach highlights the microscopic nature of the low-energy enhancement of the gSF, the so-called upbend resonance, which arises from strong particle-particle ($pp$) and hole-hole ($hh$) excitations that emerge only at finite temperature, thereby further reinsisting on the invalidity of the Brink-Axel hypothesis in this low-energy region. Overall, our study reopens the question on the long-standing problem of $^{60}$Fe production in massive stars.
\end{abstract}

\maketitle

Nucleosynthesis in stellar interiors and explosive events governs the evolution of the universe’s elemental and isotopic composition, transforming primordial hydrogen and helium from the Big Bang into the diverse suite of elements observed today \cite{Diehl2021}. The detection of $^{26}$Al \cite{Mahoney1982} and $^{60}$Fe \cite{Knie2004, Binns2016}, both in interstellar environments and terrestrial archives, provided the direct evidence that nucleosynthesis remains active in the Galaxy. With half-lives of a few million years \cite{Diehl2021}, these isotopes are short-lived compared to the Galaxy’s age and thus are well-suited for constraining stellar fusion processes and interstellar transport. The production of $^{26}$Al occurs predominantly via the proton capture on $^{25}$Mg during hydrostatic burning in massive stars \cite{Iliadis2011, Woosley1990, Prantzos1996} as well as in their supernova explosions \cite{Diehl2021}, whereas $^{60}$Fe is mostly synthesized through successive neutron captures starting from stable iron isotopes in a neutron-rich environment \cite{Diehl2021, Woosley1995, Limongi2003, Limongi2013}. The $^{60}$Fe/$^{26}$Al isotopic ratio, obtained from $\gamma$-ray measurements, has emerged as a critical probe of nucleosynthesis in massive stars, a popular constraint in the origin of solar system study and the recently recognized challenge of supernova explodability, i.e. determining which stars end their lives in supernova explosions and directly collapse into black holes \cite{Diehl2021}. The $^{60}$Fe/$^{26}$Al ratio has been determined quite accurately as 0.184 $\pm$ 0.042, by using 15 years of data from the SPectrometer on INTEGRAL (SPI) aboard the INTEGRAL mission \cite{Wang2020}. This galactic-scale measurement is consistent with the values inferred from terrestrial deposits \cite{Feige2018}. On the other hand, astrophysical models predict a wide span of ratio values, which most massive star models overestimate by a factor of 3--10, keeping its precise value still unresolved \cite{Diehl2021, Woosley2007, Limongi2006, Timmes1995, Sukhbold2016, Austin2017}. \par
Basically, theoretical predictions of the $^{60}$Fe/$^{26}$Al ratio are highly sensitive to stellar parameters \cite{Diehl2021}. Rotationally induced mixing enhances this ratio by expanding the carbon-burning shell, thereby increasing the neutron flux and promoting the $^{60}$Fe synthesis via neutron-capture processes. While both isotopes are produced in the deep interiors of massive stars, $^{26}$Al is ejected through both stellar winds and supernova explosions, whereas $^{60}$Fe is exclusively released during the supernova phase \cite{Diehl2021}. As a result, their nucleosynthetic origins trace distinct stellar regions, necessitating models that capture the full range of stellar environments to reproduce the observed $^{60}$Fe/$^{26}$Al ratio. The yields also depend sensitively on stellar mass and metallicity \cite{Limongi2006, Jones2019}. However, a major source of uncertainty stems from nuclear reaction rates \cite{Woosley2007, Limongi2006, Gao2021}, particularly the $^{59}$Fe$(n,\gamma)^{60}$Fe cross section, which dominates $^{60}$Fe production, remains experimentally unconstrained \cite{Diehl2021}. The yield scales almost linearly with this cross section in such a way that an order-of-magnitude uncertainty directly propagates to the predicted $^{60}$Fe abundance.\par
Generally, the $^{60}$Fe isotope is synthesized in astrophysical environments via the sequential neutron captures $^{58}$Fe$(n,\gamma)^{59}$Fe$(n, \gamma)^{60}$Fe. In massive stars, the $^{22}$Ne$(\alpha,n)^{25}$Mg reaction provides the dominant neutron flux. With a half-life of $\sim$2.6 Myr, $^{60}$Fe $\beta$-decays to $^{60}$Co ($T_{1/2} = 5.3$ yr), which in turn decays to excited states in $^{60}$Ni, followed by $\gamma$-ray emissions at 1173 and 1332 keV. The destruction of $^{60}$Fe occurs mainly through $^{60}$Fe$(n,\gamma)^{61}$Fe during hydrostatic shell burning, and via $^{60}$Fe$(p, n)^{60}$Co under explosive conditions. The branching at $^{59}$Fe is, therefore, crucial for the $^{60}$Fe production, making the $^{59}$Fe$(n,\gamma)^{60}$Fe reaction cross section critical. \par
To date, the experimental measurement of $^{59}$Fe$(n,\gamma)^{60}$Fe cross section and its Maxwellian average (MACS) remains a challenging task due to the short lifetime of $^{59}$Fe, which requires a neutron source with intensive flux. The first attempt to determine the $^{59}$Fe$(n,\gamma)^{60}$Fe cross section at astrophysical energies was reported in Ref. \cite{Uberseder2014}. This study was based on indirect Coulomb-dissociation technique, in which only the E1 $\gamma$-ray strength function (gSF), fitted to the measured Coulomb-dissociation cross section, was used in the statistical Hauser-Feshbach (HF) model to calculate the $^{59}$Fe$(n,\gamma)^{60}$Fe cross section. A later study was attempted to determine the $^{59}$Fe$(n,\gamma)^{60}$Fe cross section by using the indirect surrogate-ratio method, namely based on the experimental measurement of the $^{59}$Fe$(n,\gamma)$/$^{57}$Fe$(n,\gamma)$ cross-section ratio ($R$) \cite{Yan2021}. Since no experimental gSF and nuclear level density (NLD) were available at that time, the authors of Ref. \cite{Yan2021} used the experimental $R$ values in the neutron energy region from 3 to 8 MeV to constrain those quantities. Two NLD and gSF models used were phenomenological back-shifted Fermi gas (BSFG) and standard Lorentzian (SLO) models, respectively. As a result, the obtained $^{59}$Fe$(n,\gamma)^{60}$Fe cross section was determined with a large uncertainty of about 14.5\% (see e.g., Fig. 3 of Ref. \cite{Yan2021}). Indeed, constraining both NLD and gSF on the experimental neutron-capture cross sections and their relevant data, such as the surrogate ratio, is not tight enough. Within the HF framework, the neutron-capture cross section is proportional to the product of NLD and gSF, thus a low gSF can be compensated by a high NLD and vice versa. In other words, multiple combinations of NLD and gSF can lead to the same neutron-capture cross section. Therefore, the $^{59}$Fe$(n,\gamma)^{60}$Fe cross section and its MACS are still in question until the gSF and NLD are directly measured.  \par
To tackle the above problem, a recent study has employed the $\beta$-Oslo method to directly extract the gSF and NLD of $^{60}$Fe from the experimental $\beta^-$-decay spectra of $^{60}$Mn \cite{Spyrou2024}. The extracted gSF and NLD are then used in the HF model to calculate the MACS and the results obtained show a strong enhancement of factors from 1.6 to 2.1 compared to two previous estimates in Refs. \cite{Uberseder2014, Yan2021}. Although the innovative findings in Ref. \cite{Spyrou2024} are significant for the prediction of $^{60}$Fe production as well as $^{60}$Fe/$^{26}$Al ratio in massive stars, some issues still remain in this study. \par
%
In the present work, we investigate the $^{59}$Fe$(n,\gamma)^{60}$Fe cross section and its MACS using the theoretical NLD and gSF obtained within the microscopic framework of the exact pairing (EP) plus the independent-particle model (IPM) for NLD and plus the phonon damping model (PDM) for the gSF. For the latter, an extended version, in which the enhancement of gSF at the low $\gamma$ energy, known as the upbend resonance (UBR), is microscopically treated. \par
%
\section{Results}
\begin{figure}[h!]
       \includegraphics[width=1.0\textwidth]{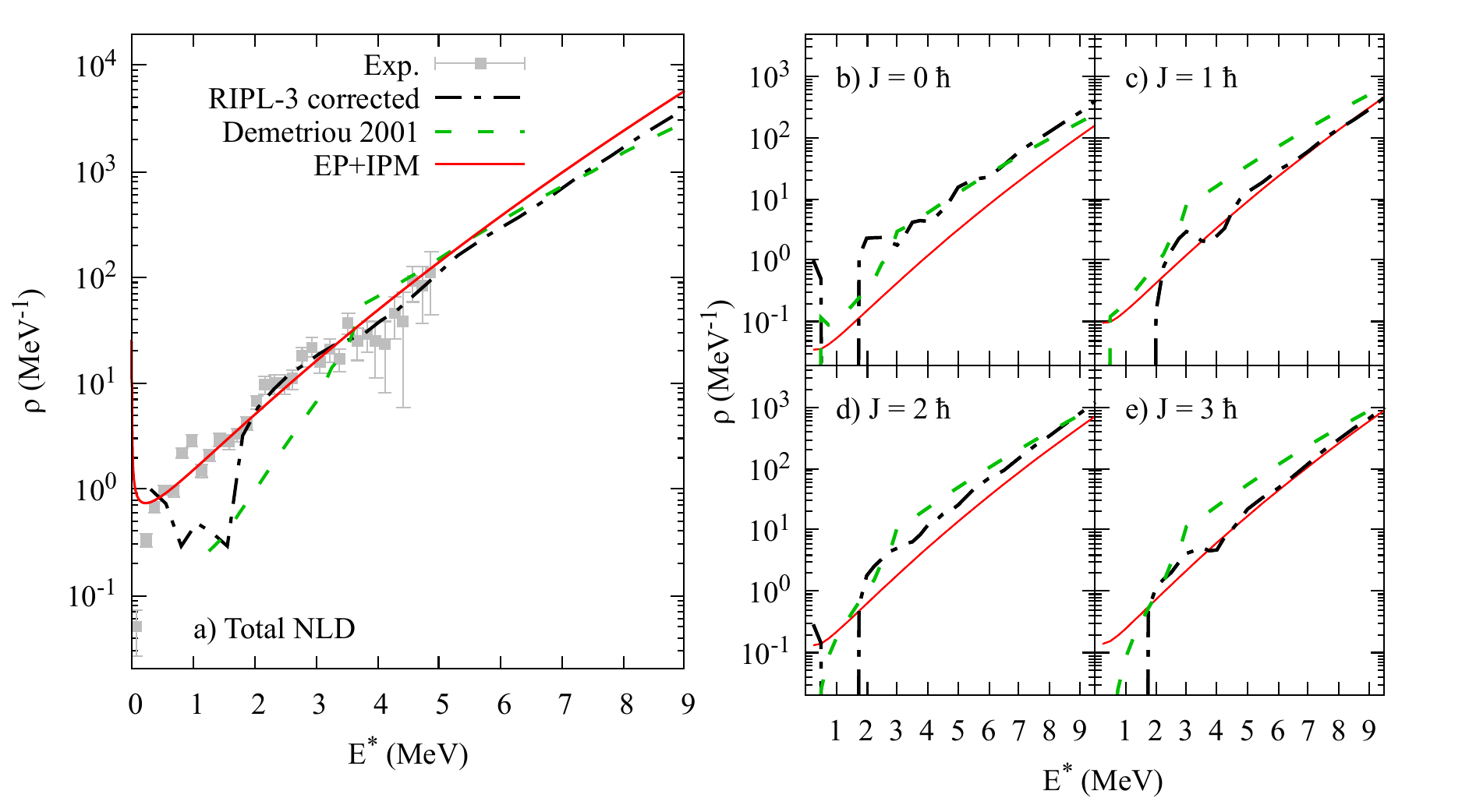}
       \caption{Total (a) and $J=0-3 ~\hbar$ [(b)$-$(e)] NLDs obtained within the EP+IPM in comparison with the experimental data extracted using the $\beta-$Oslo method \cite{Spyrou2024} as well as those predicted by two microscopic models, HFBC with readjusted parameters (RIPL-3 corrected) \cite{HFBC} and HFBCS (Demetriou 2001) \cite{HFBCS}, for $^{60}$Fe nucleus.}
        \label{Fig1}
    \end{figure}

It is seen in Fig. \ref{Fig1}(a) that the EP+IPM NLD without any normalization (solid line) nicely agrees with the experimental data (squares with error bars) in the entire excitation-energy $E^*$ range, while those predicted by two microscopic approaches, Hartree-Fock-Bogolyubov plus Combinatorial method (HFBC) with an optimal renormalization to match the measured data (RIPL-3 corrected, dash dotted line) \cite{HFBC} and Hartree-Fock plus BCS (Demetriou 2001, dashed line) \cite{HFBCS}, strongly deviate from the data at $E^*$ below $2-3$ MeV. For the angular momentum-dependent NLDs with $J=0-3 ~\hbar$, the EP+IPM with microscopic calculation of spin cut-off parameters \cite{Hung2020} predicts the NLD lower than those of the HFBC and HFBCS, particularly at $J=0 ~\hbar$ (Fig. \ref{Fig1}[(b)-(e)]). Such a difference will strongly affect the corresponding $(n,\gamma)$ cross section discussed later. We note that this is indeed the first microscopic model that provides an excellent description of the $^{60}$Fe NLD data, in particular without employing any direct fitting or adjusting parameters to the data.\par
%
%
    \begin{figure}[h!]
       \includegraphics[width=1.0\textwidth]{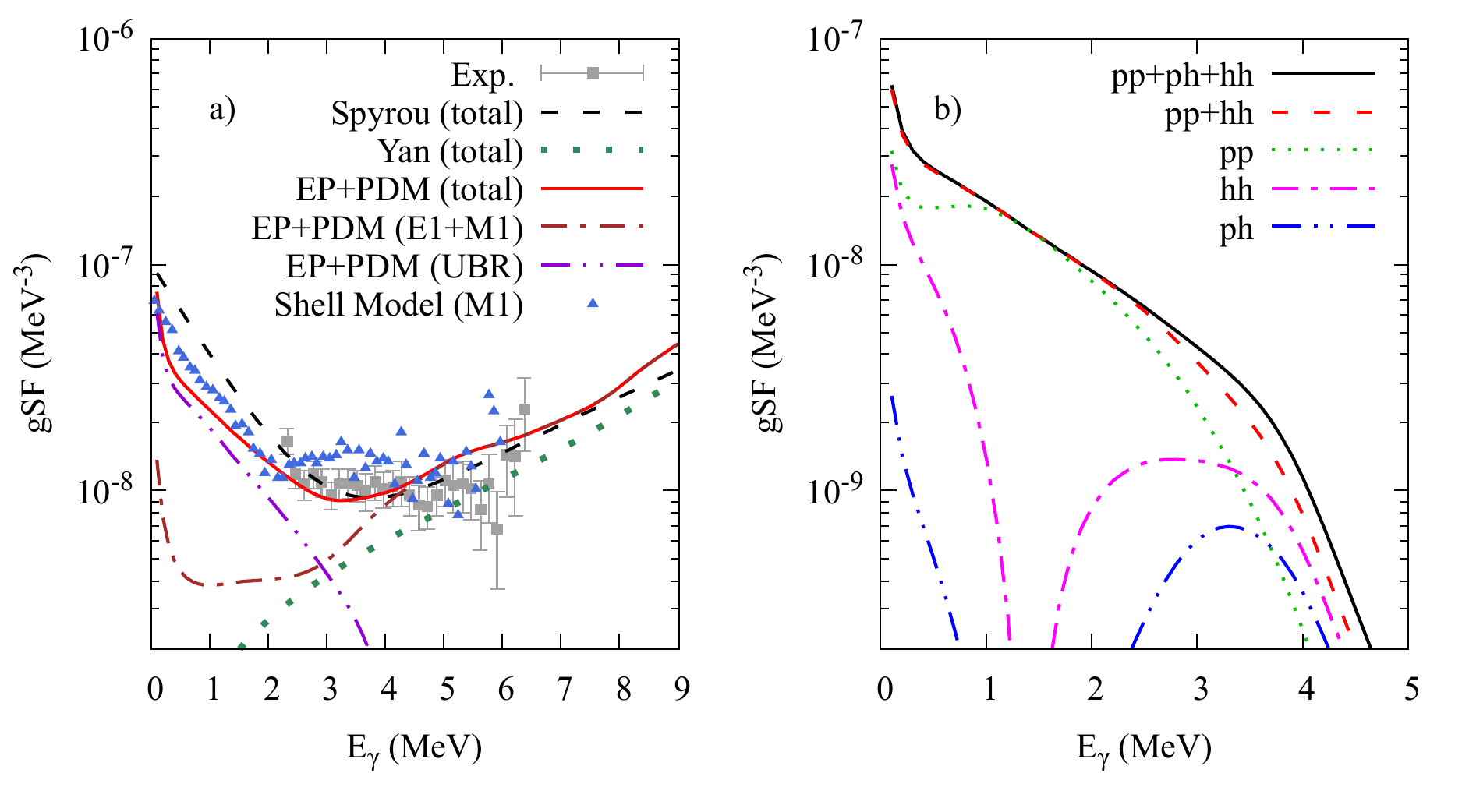}
       \caption{a) Total and partial ($E1+M1$ and UBR) gSFs obtained within the extended EP+PDM versus the experimental data extracted using the $\beta-$Oslo method \cite{Spyrou2024} as well as the phenomenological gSFs employed Yan {\it et al} [Yan (total)] and Spyrou {\it et al} [Spyrou (total)] along with the microscopic $M1$ shell-model calculation, for $^{60}$Fe.  b) Contributions of $ph$, $pp$, and $hh$ excitations to the EP+PDM UBR gSF.}
        \label{Fig2}
    \end{figure}
%
For the gSF, Fig. \ref{Fig2}a shows that the $E1+M1$ gSF obtained within the EP+PDM (dash dotted line) describes reasonably well the experimental data as well as the phenomenological gSF used in Ref. \cite{Spyrou2024} (dashed line) at $E_\gamma > 3$ MeV, whereas it significantly underestimates the data at $E_\gamma$  below 3 MeV. The total gSF employed in Ref. \cite{Yan2021} completely underestimates the experimental data. By extending the EP+PDM to include the UBR, the obtained total gSF agrees with all the data. In addition, the UBR gSF obtained within the extended EP+PDM aligns with the shell-model calculation for $M1$ mode in \cite{Schwengner2017}, thus strengthening its reliability. The UBR gSF given by Spyrous {\it et. al.} \cite{Spyrou2024}, which follows the gSF data of $^{56}$Fe (see e.g., Fig. 1(b) of Ref. \cite{Spyrou2024}), is slightly higher than those predicted by the shell model and extended EP+PDM. Such a difference is understandable as the gSF of two nuclei, $^{56}$Fe and $^{60}$Fe, might be close to each other but not the same. In Fig. \ref{Fig2}(b), we show that the UBR gSF at $E_\gamma < 4$ MeV is mainly contributed by the $pp$ and $hh$ excitations (dashed line). Notably, the $pp$ contribution (dotted line) is significantly stronger than the $hh$ one (dash dotted line). The $ph$ contribution is only substantial at $E_\gamma>4$ MeV. The fact that these $pp$ and $hh$ couplings only exist at $T>0$ invalidates the Brink-Axel hypothesis \cite{BAH} in such a low-energy regime, which assumes that the gSF is independent from the excitation energy or temperature.\par
\begin{figure}
       \includegraphics[width=1.0\textwidth]{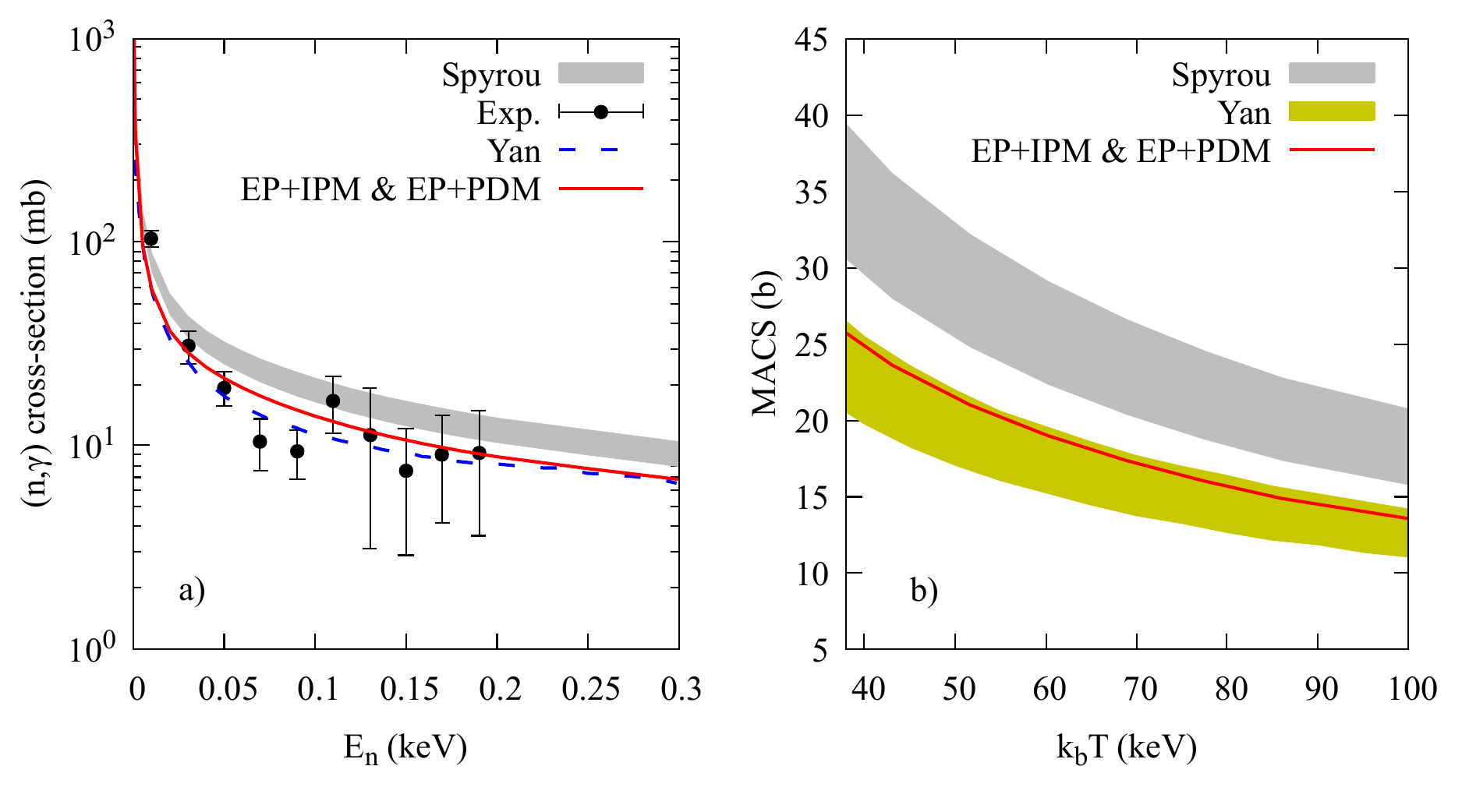}
       \caption{Comparison between cross section of $^{59}$Fe$(n,\gamma)^{60}$Fe (a) and its MACS (b) obtained within the EP+IPM \& EP+PDM with those obtained within the two most recent evaluations by Yan {\it et al} \cite{Yan2021} and Spyrou {\it et al} \cite{Spyrou2024}. The uncertainty is shown as a shaded region. The Spyrou's cross section of $^{59}$Fe$(n,\gamma)^{60}$Fe in (a) is calculated using the NLD and gSF employed to determine the MACS in (b).}
        \label{Fig3}
    \end{figure}

The EP+IPM NLD and EP+PDM gSF, which nicely match the experimental data, are then input to Talys code (v1.95) to calculate the $^{59}$Fe$(n,\gamma)^{60}$Fe cross section and its MACS within the HF statistical model framework. The obtained results are shown in Fig. \ref{Fig3} in comparison with the two most recent evaluations Refs. \cite{Yan2021,Spyrou2024}. For ease of reference, the cross sections and MACS calculated in Refs. \cite{Yan2021,Spyrou2024} are hereafter referred to as the Yan and Spyrou data, respectively. For the cross section, Fig. \ref{Fig3}(a) shows that the EP+IPM \& EP+PDM and Yan calculations agree well with the experimental data, also given in the work of Yan \textit{et al.}, whereas the Spyrou one overestimates the data. For the MACS, the EP+IPM \& EP+PDM calculation lie in the upper boundary of the Yan evaluation, which is notably lower than that of Spyrou, as shown in Fig. \ref{Fig3}(b). \par

\section{Discussion}
At the time of Yan's work, no experimental NLD and gSF were available for $^{60}$Fe. As a result, Yan {\it et al.} indirectly constrained the NLD and gSF by fitting the experimental ratio of $^{59}$Fe/$^{58}$Fe (n,$\gamma$) cross sections in the neutron energy range from 3 to 8 MeV (the surrogate method). These constrained NLD and gSF, which successfully reproduce the neutron-capture cross section of $^{59}$Fe in the low-energy $E_n = 0 - 0.3$ MeV [dashed line in Fig. \ref{Fig3}(a)], were subsequently used to calculate the MACS [yellow band in Fig. \ref{Fig3}(b)]. However, the constraints were insufficiently stringent, as the gSF employed by Yan  {\it et al.} [dotted line in Fig. \ref{Fig2}(a)] fails to describe the later experimental data by Spyrou {\it et al.}, particularly in the UBR region. \par
In contrast, Spyrou {\it et al.} directly constrain both NLD and gSF to the experimental data before employing them to determine the MACS of $^{59}$Fe$(n,\gamma)^{60}$Fe. Based on the comparison between various phenomenological and microscopic predictions and the experimental NLD data, the HFBC with an optimal renormalization was best selected, as shown in Fig. \ref{Fig1}(a). As for the gSF, Spyrou {\it et al.} assume it consists of two components, $E1$ and UBR, described by the simple Lorentzian (SLO) model and a phenomenological function, respectively, with the latter being given by $f^{\rm UBR}(E_\gamma) = C \exp(-\mu E_\gamma)$. The total gSF, defined as the sum of these two components, is then fitted to the experimental data. However, since the experimental gSF data are unavailable in the UBR region [Fig. \ref{Fig2}(a)], there should be several possible UBR gSFs, e.g., strong, medium, and weak UBRs, as shown in Fig. \ref{FigS1} of Supplementary Material. The dashed line in Fig. \ref{Fig2}(a) shows only the medium UBR gSF, which follows the gSF data of $^{56}$Fe. This results in the uncertainty of the calculated cross section and MACS, as seen in Fig. \ref{Fig3} or Fig. \ref{FigS2} of Supplementary Material. Practically, the Spyrou MACS in Fig. \ref{Fig3}(b) is obtained by combining the experimental data (total NLD and total gSF) in a limited energy region with the corresponding model calculations in the energy regions where the data are unavailable. This MACS is about $1.6-2.0$ times higher than our prediction as well as Yan's value. Such a strong enhancement and the consequent enhanced production of $^{60}$Fe in star environment thus depend on the reliability of the theoretical models used (see also Supplementary Material). Figures \ref{Fig1}(a) and \ref{Fig2}(a) show that the EP+IPM NLD and EP+PDM gSF are not much different with those of RIPL-3 corrected and Spyrou (total). However, the calculated cross sections [Fig. \ref{Fig3}(a)] and MACSs [Fig. \ref{Fig3}(b)] are strongly different. Figure \ref{Fig1}(b) clearly explains the origin of this discrepancy. In this figure, as discussed above, the $J=0$ and $2 ~\hbar$ (even spin) components of the EP+IPM NLD are significantly lower than the RIPL-3 corrected, whereas the EP+IPM $J=1$ and $3 ~\hbar$ (odd spin) NLDs are almost identical to those of RIPL-3 corrected\footnote{Our test indicates that the low-spin components of the NLD with $J \leq 3 ~\hbar$ are mainly contributed to the calculation of $(n,\gamma)^{60}$Fe cross section within the HF framework.}. Therefore, the discrepancy between two MACS values is mainly originated from the partial even-spin NLDs. It is worthwhile to mention that the $J=2 ~\hbar$ NLD is related to the $2^+$ excited states, among which the first one is strongly influenced by the pairing correlation. Within the EP+IPM, the thermal pairing is exactly treated, while it is approximated within the HFBCS (RIPL-3 corrected) approach. Meanwhile, the $J=0 ~\hbar$ NLD is associated with the monopole interaction, which forms a component of vibrational enhancement factor $k_{\rm vib}$. The latter is microscopically calculated within the EP+IPM following the statistical thermodynamic approach discussed in Ref. \cite{Hung2020}, instead of using our previous empirical formula as in Ref. \cite{Hung2017}. 
%
%
In short, the enhancement of MACS and $^{60}$Fe production in massive stars reported by Spyrou {\it et al.} in Ref. \cite{Spyrou2024} originates from the authors' choice of the employed NLD model, namely the renormalized HFBC. Although the latter is constrained to the experimental total NLD data, it remains insufficiently detailed to reliably predict the MACS, since the HF calculations require $J-$dependent NLDs. Furthermore, the $(n,\gamma)^{60}$Fe cross section predicted by employing the combination of NLD and gSF adopted by Spyrou {\it et al.}  overestimates the experimental data, as clearly seen in Fig. \ref{Fig3}(a). In contrast, based on the fully microscopic nature, the EP+IPM \& EP+PDM calculations naturally match the $(n,\gamma)$ cross section and the upper bound of MACS data in Yan {\it et al.}, thereby casting doubt on the conclusion on the enhanced MACS and $^{60}$Fe production in massive stars reported by Spyrou {\it et al}. Notably, the EP+IPM \& EP+PDM presented in the present work are the first microscopic calculations that are able to simultaneously well describe three experimental observables (NLD, gSF, and $(n,\gamma)$ cross section) for an important $^{60}$Fe nucleus. \par
To summarize, the present study shows that the MACS enhancement recently reported by Spyrou {\it et al.} \cite{Spyrou2024} stems from their use of the normalized HFBC NLD, which, despite being constrained to the total NLD data, lacks the reliability in the spin dependence for accurate HF calculations. In contrast, microscopic calculations within the EP+IPM \& EP+PDM, which simultaneously describe well, for the first time, three experimental observables, predict a significantly lower MACS. This  not only challenges the conclusion of enhanced $^{60}$Fe production in massive stars drawn by Spyrou {\it et al.}, but also highlights the critical role of microscopic nuclear structure inputs in astrophysical modeling. Besides, the microscopic nature of the UBR, which causes a considerable enhancement in the low $E_\gamma$ region of the gSF, is also explained due to the $pp$ and $hh$ excitations emerged only at finite temperature, thereby invalidating the well-known Brink-Axel hypothesis in such a low $E_\gamma$ region. Last but not least, the EP+IPM \& EP+PDM should be considered as reliable microscopic models for two nuclear physics inputs (NLD and gSF), which are essential for modeling astrophysical reaction cross sections. \par
\section{Method}
Given that the EP+IPM and EP+PDM formalisms are well-established (see e.g., Refs. \cite{Hung2017,Hung2020,Phuc2020,Le2022}), only a brief overview, tailored to the specific case of $^{60}$Fe, is presented here. Within the EP+IPM, the NLD is calculated using the canonical-ensemble partition function built on the exact solutions of the pairing Hamiltonian for a truncated set of single-particle states around the Fermi level \cite{Hung2009}, combining with that of finite-temperature independent-particle model accounted for the remaining single-particle states beyond the truncated space \cite{Alhassid2003} (for details, see Eq. (3) of Ref. \cite{Le2022}). The EP+IPM partition function, $Z_{\rm EP+IPM}$, serves as the foundation for calculating heat capacity $C$, entropy $S$, and excitation energy $E^*$ as a function of temperature $T$. 
Through the inverse Laplace transformation of $Z_{\rm EP+IPM}$, the total NLD as a function of $E^*$ is computed as $\rho(E^*) = k_{\rm vib}k_{\rm rot} \frac{\omega(E^*)}{\sigma_{\parallel}\sqrt{2\pi}}$, where $\omega(E^*)=\exp[{S}/(T\sqrt{2\pi C})]$ and $\sigma_{\parallel}$ are the total state density and the spin cut-off parameter along the symmetry axis, respectively. The latter is microscopically calculated as $\sigma_{\parallel}^2 = \frac{1}{2}\sum_k m_k^2 {\rm sech}^2(\frac{1}{2} E_k/T)$ \cite{Hung2020}, with $m_k$ and $E_k$ being the single-particle angular-momentum projections written in the deformed basis and the quasiparticle energies including the exact thermal pairing, respectively. The factors $k_{\rm vib}$ and $k_{\rm rot}$, which, respectively, account for collective vibration and rotational enhancements, are empirically calculated using Eq. (6) of Ref. \cite{Le2022}. The spin-dependent NLDs $\rho(E^*,J)$, which are essential for HF calculations, are obtained by multiplying the total NLD with the spin distribution, i.e., $\rho(E^*,J) = \rho(E^*) \frac{2J+1}{2\sigma^2}\exp{\frac{-J(J+1)}{2\sigma^2}}$, where $\sigma$ is the average spin cut-off given as $\sigma^2 = \frac{1}{2}(\sigma_\parallel^2 + \sigma_\perp^2)$. The $\sigma_\perp^2$ is related to the microscopic $\sigma_\parallel$ by the semi-empirical formula $\sigma_\perp = \sigma_\parallel \sqrt{\frac{3+\beta_2^2}{3-2\beta^2_2}}$, with $\beta_2$ being the quadrupole deformation parameter. For $^{60}$Fe, the experimental value of $\beta_2=0.224$,  estimated from the experimental B(E2) value,  is chosen according to Nudat3.0 database \cite{nudat3}.\par
Within the EP+PDM, the gSF for a given resonance is calculated as $f_{\rm R}(E_\gamma,T) =  \frac{1}{(2L+1)\pi^2\hbar^2c^2}\times\frac{\pi}{2}\frac{\sigma_{\rm  R}\Gamma_{\rm  R}(E_\gamma,T)S_{\rm  R}(E_\gamma,T)}{E_\gamma}$, where $S_{\rm  R}(E_\gamma,T)$, $\sigma_{\rm  R}$, $E_{\rm  R}$, and $\Gamma_{\rm  R}(E_\gamma,T)$ denote the strength function, cross section, energy, and width of the resonance considered, respectively. The resonance width evolves with temperature as 
$\Gamma_{\rm R}(E_\gamma,T) = 2\gamma_{\rm R}(E_\gamma,T)$, where $\gamma_{\rm R}(E_\gamma,T)$ is the resonance phonon damping computed within the EP+PDM \cite{Dang2012,Dang1998}. This damping is microscopically derived by using the quasiparticle-occupation number $n_k$ and $E_k$ obtained within the EP. The strength function $S_{\rm  R}(E_\gamma,T)$ takes the form as $S_{\rm R}(E_\gamma, T) = \frac{1}{\pi} \frac{\gamma_{\rm  R}(E_\gamma, T)}{(E_\gamma - E_{\rm  R})^2 + [\gamma_{\rm  R}(E_\gamma, T)]^2}$ \cite{Phuc2020,Hung2017}. For the electric dipole $E1$ resonance,  the EP+PDM introduces two parameters: $F^{\rm R}_1$ and $F^{\rm R}_2$. $F^{\rm R}_1$ accounts for all the couplings of the resonance phonon to the collective particle-hole ($ph$) excitations and is adjusted to reproduce the giant-dipole resonance (GDR) width at $T=0$. $F^{\rm R}_2$ is responsible for those in associated with the non-collective particle-particle ($pp$) and hole-hole ($hh$) excitations, and is chosen at $T=0$ so that the GDR energy remains nearly constant with increasing temperature \cite{Hung2017,Dang1998}. For $^{60}$Fe, the total gSF is mainly contributed by the $E1$ and $M1$ (magnetic dipole) resonances, along with the UBR, namely $f_{\rm EP+PDM} = f_{E1}^{\rm I} + f_{E1}^{\rm II} + f_{M1} + f_{\rm UBR}$\footnote{Since $^{60}$Fe is deformed, its GDR is split into two E1 resonances with energies $E1^{\rm I}$ and $E1^{\rm II}$.}. The resonance parameters involved in the EP+PDM calculation are provided in Table \ref{tab1}. For $E1$ components, all the parameters are taken from the global parameterization of Steinwedel-Jensen \cite{IAEA1506}, with slight adjustments to $E1^{\rm I}$ and $E1^{\rm II}$. The $M1$ parameters are also calculated using the global parameterization of spin-flip giant resonance \cite{IAEA1034}. The UBR is microscopically treated following the extended EP+PDM framework, which has been recently developed \cite{PhucUBR} based on a hypothesis that it is originated from the interaction of $p/h$ states with a collective excitation at the very-low $E_\gamma$. Within the conventional PDM, only $ph$ configurations are present at $T= 0$, whereas $pp$ and $hh$ excitations appear solely at $T > 0$, whose couplings to the resonance phonon causes the damping of the latter. In the UBR region ($E_\gamma < 3$ MeV), there are few or even no $ph$ couplings with sufficiently low energy to generate transitions. The UBR thus cannot develop at $T = 0$. However, at finite temperatures, $pp$ and $hh$ excitations begin to contribute in the $E_\gamma < 3$ MeV region, enabling the emergence of the UBR. In particular, the strength of these $pp$ and $hh$ excitations is approximately three times that of the GDR as reported in Ref. \cite{PhucUBR}. Regarding the UBR parameters, the $E^{\rm UBR}$ value can be estimated based on the empirical M1 gSF extracted from discrete level schemes using the method described in Ref. \cite{Midtbo2018}, while its cross section $\sigma^{\rm UBR}$ is determined so that the ratio between the integrated strength of the gSF in the UBR region to that of the total gSF and the mass number follows the global relation found in Ref. \cite{PhucUBR}, namely $R_{\rm fit}=1.03$ for $^{60}$Fe. As for the UBR width, its value (4.08 MeV) is not shown in Table \ref{tab1} as it is directly calculated based on all the $pp$, $hh$ and $ph$ excitations, whose coupling strengths are approximately three times those of the GDR. Meanwhile, the nuclear temperature was chosen at $T=1.1$ MeV, in agreement with the $T\sim 1.0$ MeV value predicted using the discrete nuclear levels as well as the constant temperature model \cite{RIPL-3,HFBC}. The selection of $E1, M1$, and UBR parameters in Table \ref{tab1} preserves the microscopic nature of the extended EP+PDM employed in the present study, i.e., no direct fitting parameter to the experimental gSF data is used. \par
%
\begin{table}
\centering
\caption{\label{tab1} Resonance parameters employed in the extended EP+PDM calculation for $^{60}$Fe nucleus.}
\begin{tabular}{cccc}
\hline \hline
Resonance ($\rm R$) & \textbf{$E^{\rm R}$ } & \textbf{$\Gamma^{\rm R}$ $(T=0)$} & \textbf{$\sigma^{\rm R}$} \\
                                & (MeV) & (MeV) & (mb) \\
\hline
$E1^{\rm I}$ & 17.2 & 5.35 & 33.31 \\
$E1^{\rm II}$ & 20.3 & 7.67 & 66.63 \\
$M1$ & 10.5 & 4.0 & 2.02 \\
UBR & 1.4 & - & 0.35 \\
\hline \hline
\end{tabular}
\end{table}
\section*{Data availability}
Source data are provided with this paper in the Source Data file.
\section*{References}
%

\acknowledgments
We would like to thank Prof. A. Spyrou for providing the total gSF [dashed line in Fig. \ref{Fig2}(a)] and valuable discussions regarding our reproduced calculations of the gSF and MACS in Figs. \ref{FigS1} and \ref{FigS2}. N.N.A. gratefully acknowledges the support of Phenikaa University.
\section*{Author contributions}
N. Q. Hung performed the calculation of EP+IPM NLD and prepared the draft of the manuscript. L. T. Phuc performed the calculation of EP+PDM gSF. S. Lakshan and N. N. Anh performed the calculations of (n,$\gamma$) cross-section and MACS. N. N. Anh and N. Q. Hung performed the data analysis and data visualization. B. Dey and D. Pandit took care of data validation and contributed to prepare the draft of the manuscript. S. Bhattacharya, L. T. Q. Huong, and N. Dinh Dang contributed to manuscript preparation. All the authors discussed and interpreted the numerical data.
\section*{Competing interests}
The authors declare no competing interests.
\section*{Additional information}
\textbf{Correspondence} requests for materials should be addressed to N. Q. Hung, N. N. Anh, and B. Dey. 

\clearpage
\section*{supplemental material}
\setcounter{figure}{0}
\renewcommand{\thefigure}{S\arabic{figure}}
Figure \ref{FigS1} shows that the Spyrou's MACS range (Fig. 1(a) in Ref. \cite{Spyrou2024}) can be well reproduced by using Spyrou's $E1$ gSF, RIPL-3 (HFBC) corrected NLD combined with three strong, medium, and weak UBR gSFs [Fig. \ref{FigS1}(a)]. Specifically, the upper, middle and lower bounds of the MACS range are obtained using the RIPL-3 corrected NLD combined, respectively, with the strong, medium, and weak UBR gSFs [Fig. \ref{FigS1}(b)]. This means that the uncertainty in the MACS reported in Ref. \cite{Spyrou2024}) comes from the UBR gSFs, not the $E1$ gSF and/or the NLD models. Dotted line in Fig. \ref{FigS1}(b) confirm our assessment, namely the MACS calculated using Spyrou's $E1$ + medium UBR gSF and Demetriou 2001 (HFBCS) NLD is significantly enhanced compared to Spyrou's MACS range, although the HFBCS total NLD does not strongly deviate from those of the EP+PDM or RIPL-3 corrected. Therefore, the calculated MACS is very sensitive to the $J$-dependent NLDs, in which the even spin components have crucial contributions. This is further confirmed by the MACS values obtained using different combinations of NLD and gSF in Fig. \ref{FigS2}. For instance, if the Spyrou NLD and EP+PDM gSF is used, we obtain the MACS (dash dotted line) falling in the MACS range. Meanwhile, if the EM+IPM NLD and Spyrou gSF is used, the obtained MACS lies between the MACS ranges of Spyrou and Yan.
%
    \begin{figure}[h]
       \includegraphics[width=1.0\textwidth]{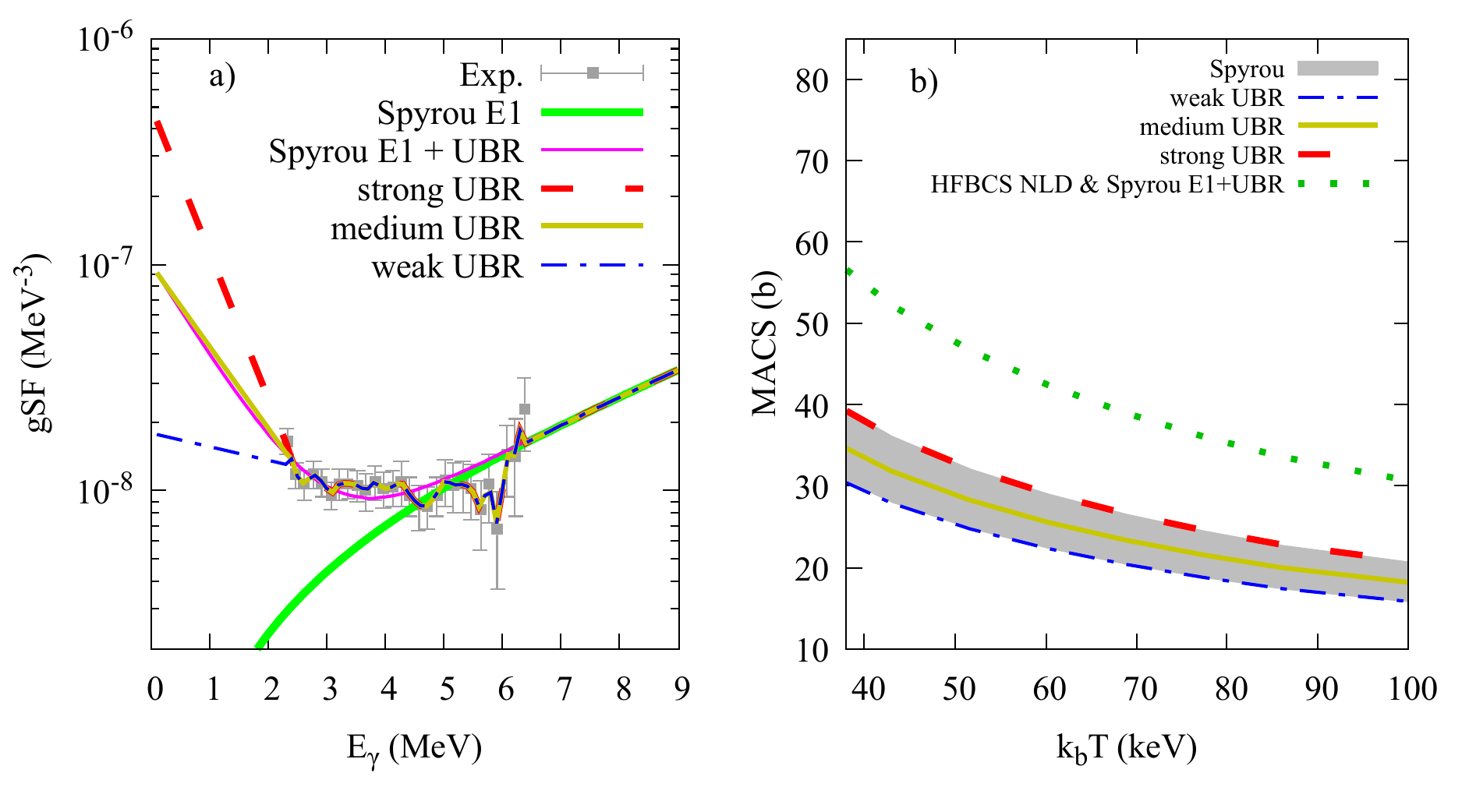}
       \caption{(a) Total and partial ($E1$, UBR and $E1$+UBR) gSFs employed in Spyrous {\it et al} \cite{Spyrou2024}. (b) Reproduction of MACS range (Fig. 1(a) in \cite{Spyrou2024}) using Spyrou's $E1$ gSF combined with three strong, medium, and weak UBR gSFs in (a) and RIPL-3 corrected NLD. Dotted line is the MACS obtained using HFBCS (Demetriou 2001) NLD and Spyrou's $E1$ + medium UBR gSF.}
        \label{FigS1}
    \end{figure}
    \begin{figure}[h]
       \includegraphics[width=0.5\textwidth]{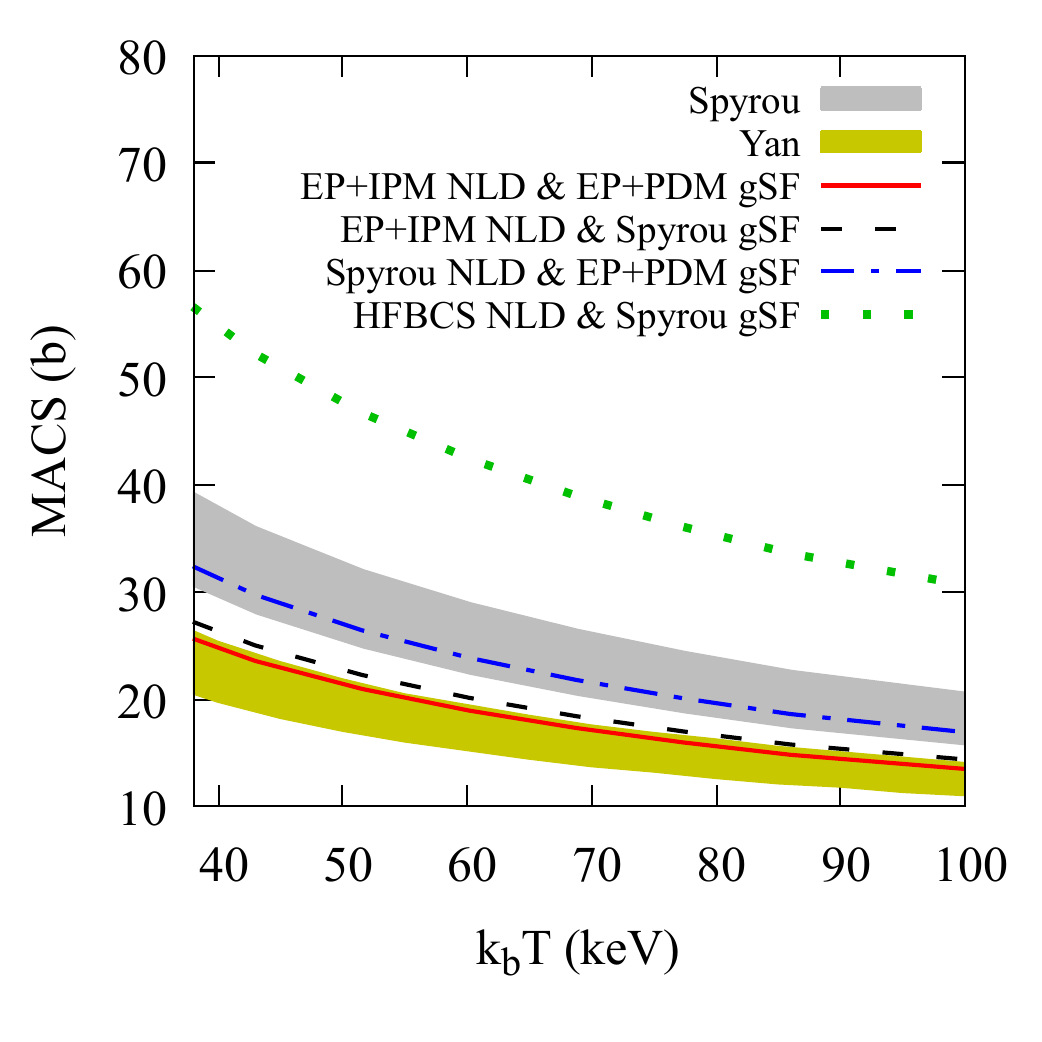}
       \caption{The MACSs obtained using different combinations of NLD and gSF.}
        \label{FigS2}
    \end{figure}
\end{document}